\documentclass[12pt,a4paper]{conference}

\usepackage{fancyhdr}
\usepackage{graphicx,amsmath,amssymb,cite}
\usepackage{multind}
\makeindex{author} \makeindex{subject}
\newcommand{\beq}{\begin{equation}}
\newcommand{\eeq}[1]{\label{#1}\end{equation}}
\newcommand{\eeqn}{\end{equation}}

\newcommand{\beqa}{\begin{eqnarray}}
\newcommand{\eeqa}[1]{\label{#1}\end{eqnarray}}
\newcommand{\eeqan}{\end{eqnarray}}

\let\bar=\overbar

\newcommand{\Dslash}{\not{\hbox{\kern-4pt $D$}}}
\newcommand{\dslash}{\not{\hbox{\kern-2pt $\del$}}}

\newcommand{\msb}{{\bar{\ssstyle M \kern -1pt S}}}

\begin{document}

\Chapter{$\pi N\rightarrow \eta N$ process in a $\chi$QM approach}
     {$\pi N\rightarrow \eta N$ process in a $\chi$QM approach}{J. He \it{et al.}}
\vspace{-6 cm}\includegraphics[width=6 cm]{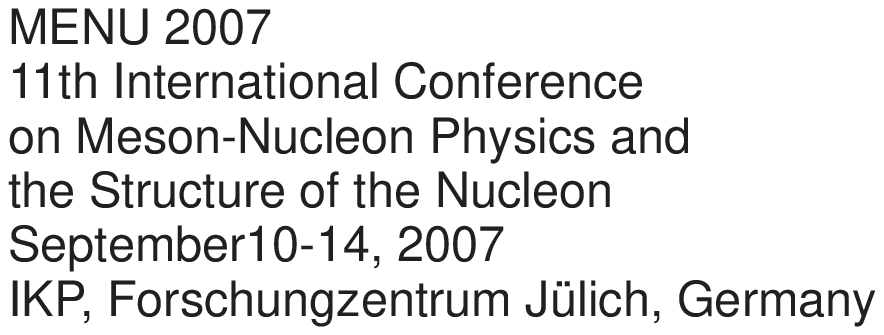}
\vspace{4 cm}

\addcontentsline{toc}{chapter}{{\it N. Author}} \label{authorStart}

\begin{raggedright}

Jun He$^{\star}$, Xianhui Zhong$^{+}$, Bijan Saghai$^{\star}$, Qiang Zhao$^{+}$
\bigskip\bigskip

$^{\star}$Laboratoire de recherche sur les lois fondamentales de l'Univers,\\
DSM/DAPNIA/SPhN, CEA/Saclay, 91191 Gif-sur-Yvette, France\\
$^{+}$Institute of High Energy Physics, Chinese Academy of Sciences,
Beijing, 100049, P.R. China\\

\end{raggedright}

\begin{center}
\textbf{Abstract}
\end{center}
A chiral quark model approach is used to
investigate the $\pi^{-}p \rightarrow \eta n$ process at low
energies.  The roles of the most relevant nucleon resonances in $n\leq 2$
shells are briefly discussed.

\section{Introduction}

The $\pi^{-}p \rightarrow \eta n$ reaction
provides a suitable probe to investigate the structure of low-lying
nucleon resonances as well as the $\eta N$ interaction.

Recent high precision data released by the BNL Crystal Ball Collaboration~\cite{CB}
has revived the interest in that process. The impact of those data on
the meson-baryon interactions has been emphasized by the SAID Group~\cite{SAID}.
Extensive theoretical efforts are also being deployed {\it via}
coupled-channel formalisms, such as the
K-matrix approach~\cite{Geisen}, meson-exchange model~\cite{Juelich},
chiral model~\cite{Valencia}, T-matrix~\cite{Zagreb},
and dynamical formalism~\cite{EBAC}.

We have extended to the $\pi N\rightarrow \eta N$ process
a comprehensive and unified approach~\cite{Li} to the meson photoproduction,
based on the low energy QCD Lagrangian in terms of quarks degrees of freedom.
This latter formalism has been
developed and proven~\cite{Sag} to be successful in investigating
$\gamma p\rightarrow \eta p, K^+ \Lambda$ and $\gamma N\rightarrow
\pi N$ reactions. In this approach, only a few parameters are required. In
particular, only one parameter is needed for the nucleon resonances to be
coupled to the pseudoscalar mesons. All the resonances can be
treated consistently in the quark model.

\section{Theoretical frame}

In the chiral quark model, the low energy quark-meson interactions
are described by the effective Lagrangian
\begin{eqnarray} \label{lg}
\mathcal{L}=\bar{\psi}[\gamma_{\mu}(i\partial^{\mu}+V^{\mu}+\gamma_5A^{\mu})-m]\psi
+\cdot\cdot\cdot,
\end{eqnarray}
\noindent where vector ($V^{\mu}$) and axial ($A^{\mu}$) currents
read
\begin{eqnarray}
V^{\mu} =
 \frac{1}{2}(\xi\partial^{\mu}\xi^{\dag}+\xi^{\dag}\partial^{\mu}\xi)~,~A^{\mu}=
 \frac{1}{2i}(\xi\partial^{\mu}\xi^{\dag}-\xi^{\dag}\partial^{\mu}\xi),
\end{eqnarray}
with $ \xi=\exp{(i \phi_m/f_m)}$, where $f_m$ is the meson decay
constant. $\psi$ and $\phi_m$ are the pion and quark fields,
respectively.

The $\eta$ meson production amplitude can be expressed in terms of
Mandelstam variables,
$\mathcal{M}=\mathcal{M}_{s}+\mathcal{M}_{u}+\mathcal{M}_t$.

The {\it s-} and {\it u-}channel transitions are given by:
\begin{eqnarray}
\mathcal{M}_{s}=\sum_j\langle N_f |H_{\eta} |N_j\rangle\langle N_j
|\frac{1}{E_i+\omega_\pi-E_j}H_{\pi }|N_i\rangle, \label{sc}\\
\mathcal{M}_{u}=\sum_j\langle N_f |H_{\pi }
\frac{1}{E_i-\omega_\eta-E_j}|N_j\rangle\langle N_j | H_{\eta}
|N_i\rangle, \label{uc}
\end{eqnarray}
\noindent where $\omega_\pi$ and $\omega_\eta$ are the energies of
the incoming $\pi$-meson and outgoing $\eta$-meson, respectively.
$H_\pi$ and $H_\eta$ are the standard quark-meson couplings at tree
level. $|N_i\rangle$, $|N_j\rangle$, and $|N_f\rangle$ stand for the
initial, intermediate, and final state baryons, respectively, and
their corresponding kinetic energies are $E_i$, $E_j$, and $E_f$.

Given that the $a_0$ meson decay is dominated by $\pi \eta$
channel~\cite{PDG}, we consider the $a_0$ exchange as the prominent
contribution to the {\it t-}channel,
\begin{eqnarray} \label{tch}
\mathcal{M}_t=\sum_j {\frac{g_{a_0\pi \eta } g_{a_0 qq}
m^2_\pi}{t^2-m^2_{a_0}} \langle N_f | \bar{\psi}_j\psi_j\vec{a}_0
|N_i\rangle}.
\end{eqnarray}
\noindent where  $m_{a_0}$ is the mass of the $a_0$ meson.


With above effective Lagrangian and following the procedures used in
Ref.~\cite{Li}, we obtain the amplitude in the harmonic oscillator
basis ~\cite{IHEP}.

\section{Results and discussion}

Using the formalism sketched above, we have investigated the cross-section for the
$\pi^{-}p \rightarrow \eta n$ process. In our model, non-resonant components include
nucleon pole term, {\it u-}channel contributions (treated as degenerate to the
harmonic oscillator shell $n$), and
{\it t-}channel contributions due to the $a_0$-exchange.

The resonant part embodies the following $n$=1,2 shell nucleon resonances:
\begin{itemize}
 \item $n$=1: $S_{11}(1535)$, $S_{11}(1650)$, $D_{13}(1520)$, $D_{13}(1700)$,
 and $D_{15}(1675)$,
 \item $n$=2: $P_{11}(1440)$, $P_{11}(1710)$, $P_{13}(1720)$, $P_{13}(1900)$,
$F_{15}(1680)$, and $F_{15}(2000)$.
\end{itemize}

Here we use the Breit-Wigner masses and widths given in the PDG\cite{PDG}.
For meson-nucleon-nucleon couplings we adopt $g_{\pi NN}$=13.48 and
$g_{\eta NN}$=0.81.

Our results for the differential cross-section are depicted in
Fig.~[1] for pion incident momenta $P_{\pi}^{lab}$ = 0.718,
0.850, and 1.005 GeV, corresponding to the total centre-of-mass energies W = 1.507,
1.576, and 1.674 GeV, respectively.
\begin{figure}[hb]
\begin{center}
\includegraphics[height=10 cm, width=13. cm]{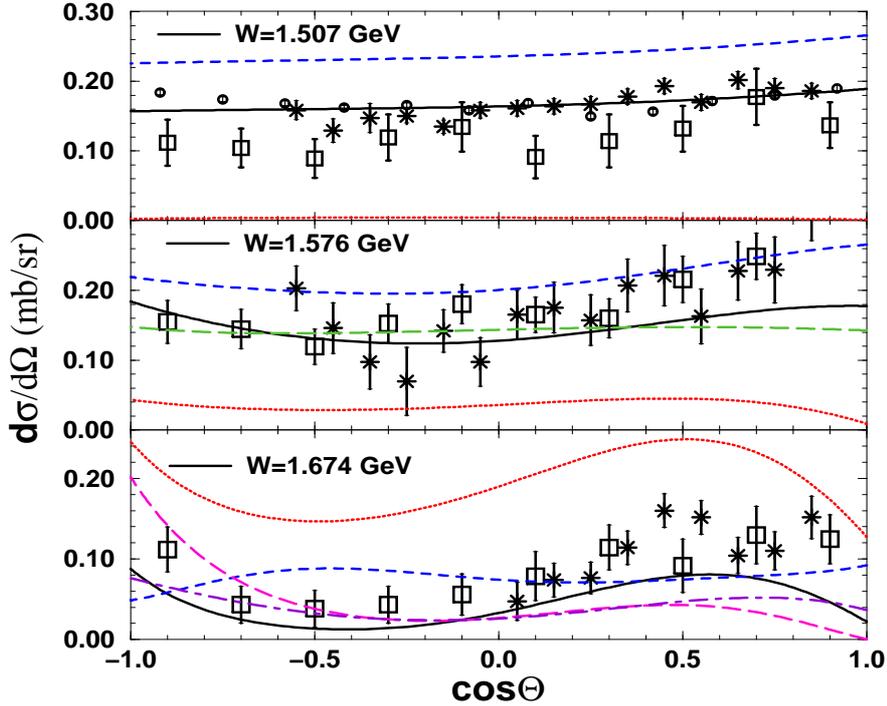}
\hspace{4. cm} \caption{Differential cross-section for $\pi^-
p\rightarrow \eta n$. The curves appearing in all the three boxes
are: full model (solid black), the $S_{11}$(1535) switched off
(dotted red), and the $S_{11}$(1650) switched off (dashed blue). In
the middle box: the $D_{13}$(1520) switched off (long dashed green).
In the lower box: the $P_{11}$(1710) switched off (long dashed
magenta) and without the $n$=2 shell contributions (dot-dashed
violet). Data are from Prakhov {\it et al.}~\cite{CB} (circles),
Richards {\it et al.}~\cite{Richards} (squares), and Deinet {\it et
al.}~\cite{Deinet} (stars).} \label{fig:Fig1}
\end{center}
\end{figure}

We get a good agreement with the data at those energies (full curves).
In order to single out the importance of various resonances, at each energy
we show results while one {\it significant} resonance is switched off.
The $S_{11}(1535)$ plays a crucial role in this energy range. At the lowest energies
it has a constructive effect, while at the highest one its contribution
becomes destructive. The $S_{11}(1650)$ has a (much) smaller and destructive effect.
The role of the $D_{13}(1520)$, shown at W=1.576 GeV, is merely to produce the
right curvature. At the highest energy, although the overall contribution from
$n$=2 shell is rather small, the $P_{11}(1710)$ produces significant effects.
This point was emphasized in our recent work~\cite{IHEP}, and led us to adopt here a
reversed sign for that resonance {\it from the beginning}. That sign change for the
$P_{11}(1710)$ could be an indication, e.g. for the breakdown of the non-relativistic
constituent quark model or for unconventional configurations inside that resonance.
More investigation is needed to underpin the origins of this novelty.



\end{document}